\documentclass[preprint,12pt,3p,onecolumn]{elsarticle}
\usepackage{blindtext}
\usepackage{fancyhdr}
\usepackage[fleqn]{amsmath}
\usepackage{amsthm,enumitem}
\usepackage{array}
\usepackage{graphicx}
\usepackage{nccmath}
\usepackage{amssymb}
\usepackage{color}
\usepackage{etoolbox}
\usepackage{algorithm}
\usepackage{epstopdf}
\usepackage[noend]{algpseudocode}
\usepackage{setspace}
\usepackage{thmtools}
\usepackage{xpatch}
\usepackage[table]{xcolor}
\usepackage{tabularx}
\usepackage{graphicx}

\usepackage[bottom]{footmisc}
\usepackage{fancyhdr}

\allowdisplaybreaks

\begin{document}
\begin{frontmatter}

\author{Hassan Yeganeh and Javane Rostampoor\\
Iran Telecommunication Research Center, Tehran, Iran\\
yeganeh@itrc.ac.ir, j.rostampoor@itrc.ac.ir }
\title{Downlink User Association and Uplink Scheduling for Energy Harvesting Users in Software-Defined Mobile Networks}

\begin{abstract}
In this paper we consider a heterogeneous network which consists of a macro base station and some pico base stations utilizing massive MIMO and MIMO techniques, respectively.
A central software-defined mobile network (SDMN) controller is adopted in order to provide user association and energy scheduling. The users are considered battery limited and are capable of simultaneous wireless information and power transfer (SWIPT) in order to harvest energy and address the energy shortage issue. These users harvest energy from the received signals in the downlink and consume it via their uplink communications.
This paper deals with the downlink user association by jointly optimizing the overall sum-rate of the network and the harvested energy by introducing an appropriate utility function. In this regard, the optimum user association and  power splitting factor for each user are calculated via the downlink optimization stage.
Then, the process of uplink scheduling is defined as choosing the best users in each time epoch to transfer data as well as optimizing their transmit power by solving Lyapunov drift-plus-penalty function. Simulation results are provided in order to confirm the optimality of the proposed algorithm in comparison with the previous user association and uplink scheduling approaches in terms of providing fairness and battery management among users.

\end{abstract}


\end{frontmatter}

\section{Introduction}\label{sec:Introduction}

Nowadays, significant enhancement in energy consumption as well as communication networks' development have led to a remarkable attention to the optimization of energy and other limited resources \cite{energy}. On the other hand, the drastic increase in the internet of things (IoT) devices results in a significant enhancement in the energy demand. In this regard, smart solutions for the wireless environment are inevitable due to the increase of the wireless communication demands and their enormous energy utilizations \cite{smart1}. Smart cities are the appropriate solutions to deal with the mentioned need since they are capable of providing energy efficiency in the networks. Energy harvesting, i.e. exploiting the energy of propagated radio frequency signals in the environment is studied in \cite{EH1,EH2,EH3}. This technique is a promising solution which is adopted in smart cities in order to tackle the energy shortage problem and prolong the life-time of low-power devices \cite{smart2}.
 In order to improve energy harvesting, simultaneous wireless information and power transfer (SWIPT) is introduced in energy harvesting capable devices  to transfer information and energy at the same time \cite{Shi,Lee,Zhao}.
In order to obtain SWIPT benefits, the receiver can exploit each of two mechanisms: time switching (TS) or power splitting (PS). The TS receiver periodically switches between energy harvester and information decoder, while the PS receiver splits the received power into harvesting power and information decoding power \cite{mine}.

 In addition, in LTE and 5G cellular networks, the introduction of heterogeneous networks (HetNets) can provide operators and users with significant benefits such as  improvement in spectral and energy efficiency as well as coverage enhancement \cite{hetnet}. On the other hand, the operators encounter difficulties by adding more complexities due to employing a heterogeneous network design \cite{hetnet2}. Consequently, the existence of a central controller is inevitable in order to manage the overall system and optimize it in order to cover the needs of users and operators. Software-defined network (SDN) is a network architecture approach to provide operators with easier management as well as control and change of the network by dividing the control plane and data plane \cite{sdn2}. In other words, SDN is capable of making smart cities a reality \cite{sdnsmart}.
  Extending SDN to wireless and mobile networks, i.e. software-defined wireless network (SDWN) and software-defined mobile network (SDMN) is a new trend of industry and academia to address nowadays needs of wireless environments \cite{sdwn1,sdwn2}. Softnet, a structure based on an SDN core network and a software-defined radio access network is introduced in order to improve the network performance as well as its efficacy \cite{sdwn3}. SDN principles in wireless cellular networks lead to a central controller which collects the reports from the base stations and users, performs some assignments and calculations and then informs the network nodes about the configurations by using the same reporting interface \cite{Arslan}.

On the other hand, multiple-input multiple-output (MIMO) technology plays a significant role in 4G and beyond generations of mobile communications especially in throughput and network efficiency enhancement \cite{mimo}. In addition, Massive MIMO has attracted a significant attention recently, since it is capable of improving the network performance \cite{Xu,massive,massive2}. Consequently, exploiting HetNets and adopting MIMO and massive MIMO techniques are definite in the next generations of wireless networks.


In \cite{Tuncer}, an SDN based placement algorithm is adopted in order to determine the allocation of managers and controllers which supports both static and dynamic resource management. The authors in \cite{Zhang} have considered a vehicular heterogeneous SDN and the ultimate goal is to minimize communication cost. The SDN role in 5G mobile networks in improving system capacity and its performance is studied in \cite{Sun}, especially in managing the heterogeneous resources. The problem of resource management over 5G-SDWN is investigated in \cite{sdwn4} where the performance of user association in SDWN is compared with the traditional user assignment and also its practical challenges and implementation requirements are discussed.

 As it is mentioned, in order to provide users and operators with their needs and utilize the benefits of HetNets as much as possible, the necessity of SDN architecture is inevitable. In this regard, an SDN controller is responsible for user management and its mobility and connection control as well as energy scheduling of battery limited users. User association, i.e. assigning the users to their optimum cells, is a critical issue in HetNets and is investigated in  \cite{Xu,Moon,Bethanabhotla}. In \cite{Xu}, the problem of user association in a HetNet is considered while the macro base station (MBS) is equipped with massive MIMO and pico base stations (PBSs) are conventional multiple-antenna nodes. The problem of user association in this paper is based on the maximization of the sum-rate of the network while adopting centralized and distributed optimization. The authors in \cite{Moon} have considered a cognitive heterogeneous network and their objective is to achieve the optimal user association along with adopting some constraints on the interference.  The problem of load balancing and user association with massive MIMO BSs is investigated in \cite{Bethanabhotla}, where the objective function is defined based on load balancing and providing the users with more fairness.

 On the other hand, Uplink scheduling is also of the great importance in wireless networks. In this regard, the authors in \cite{Goonewardena} have studied the problem of fair scheduling in multiple-antenna energy harvesting nodes which is achieved by maximizing the data-rate and regulating fairness and stabilization of stored energy in the nodes. In order to obtain these goals, the authors have employed Lyapunov drift-plus-penalty function which represents the queue nature of  batteries. The previous works in UL energy scheduling have focused on analysing resource allocation in simple transmitter-receiver environment. However, in  real scenarios we face complex cellular networks even heterogeneous ones to address next generation wireless network's needs.
  To the best of our knowledge, the problem of joint uplink (UL) and downlink (DL) scheduling in energy harvesting heterogeneous networks while considering user fairness according to the queue feature of the battery of users has not been studied in the literature before.

 In this paper, we deal with the problem of user association via downlink and battery scheduling in uplink while considering a heterogeneous network consists of an MBS equipped with massive MIMO and some MIMO PBSs. In addition, the users are suffering from battery capacity limitation and hence they are capable of energy harvesting in order to address this problem. The user association is derived based on the maximization of a utility function which is an appropriate function of total sum-rate and received energy. However, the previous works have applied user association based on the users' distance from the BS or the total sum-rate. The introduced approach provides the network with the overall utility function enhancement and a suitable balance between the data-rate and energy.  The users harvest energy in the downlink and exploit it in the uplink in order to send their information. The SDN controller, monitors the process of downlink user association and uplink battery scheduling. Managing uplink data transfer includes choosing the best user in each cell and time epoch to transmit its data and determining its transmit power in order to satisfy the operator's benefit and also fairness among users. In order to achieve this goal and obtain the optimum battery scheduling, a Lyapunov drift-plus-penalty function is employed. Our simulation results validate the efficiency of our proposed method in comparison with other conventional DL and UL scheduling in terms of providing fairness among users as well as an acceptable trade-off between data-rate and energy.

\setlength{\arrayrulewidth}{1mm}
\setlength{\tabcolsep}{30pt}
\renewcommand{\arraystretch}{1}

{\rowcolors{2}{gray!30!white!50}{gray!5!white!40}
\begin{table}[!h]
\begin{tabular}{!{\vline}p{2.1cm}!{\vline}p{11cm}!{\vline} }

\hline
Parameter & Description\\
\hline
$J$  & Total Number of Base Stations.   \\
$K$ &  Total Number of Users.    \\
${\bf h}_{j,k}$ & vector of channel coefficients between BS $j$ and user $k$.  \\
${\bf g}_{j,k}$  &Small-scale fading coefficients between BS $j$ and user $k$.  \\
${l}_{j,k}$ &  large-scale fading coefficients between BS $j$ and user $k$. \\
${\bf H}_{j}$ &  Matrix of channel coefficients at BS $j$.  \\
${\bf y}_{j}$ & Received signals by users connected to BS $j$.\\
${\bf W}_j$ & Precoding Matrix of BS $j$. \\
${\bf d}_j$ &  Data vector of BS $j$.\\
${\bf n}_j$ & Noise vector at BS $j$.\\
$R_{j,k}^{DL}$ & Downlink data-rate between BS $j$ and user $k$.\\
$x_{j,k}$ &  User association parameter.\\
$\alpha_{j,k}$ & Harvesting factor of user $k$ connected to BS $j$.\\
$M_j$ &  Number of MBS antennas.\\
$L_j$ & Maximum number of users that can be served.\\
$P_j$ &  Transmission power of BS $j$.\\
$P_k$ &  Transmission power of user $k$.\\
$\sigma_k^2$ &  Noise power at user $k$.\\
$\sigma_j^2$ &  Noise power at BS $j$.\\
${{\sigma}_{D_k} ^2}$ &  Additional noise power of decoding at user $k$.\\
$RP_{j,k}$ &  Received power at user $k$ from BS $j$.\\
$U_{j,k}$ &   Utility function of user $k$ connected to BS $j$.\\
$R_{j,k}^{UL}$ &  Uplink data-rate between BS $j$ and user $k$.\\
$\lambda_{j,k}$ &  Eigenvalue of channel between user $k$ and BS $j$.\\
$S_{k}$ &  Harvested power at user $k$.\\
$C_{k}$ &  Maximum battery capacity of user $k$.\\
$V_{k}$ &  Control Parameter.\\
$t_{coh}$ &  Average coherence time of the channels.\\
$t_{end}$ &  Time of end of the process.\\

\hline
\end{tabular}
\caption{List of all parameters throughout the paper.}
\end{table}
}


Rest of this paper is organized as follows. In Section 2, the proposed system model is introduced. The optimization problem is formulated and tackled in Section 3 and the control parameter is discussed in Section 4. The simulation results are proposed in Section 5 and finally the paper is concluded in Section 6.

In the paper, lower-case letters and bold lower-case letters represent scalar variables and vectors, respectively. Matrices are denoted by bold upper-case letters.
 $\left( . \right) ^{H}$ stands for the complex conjugate transpose and $|.|$ is the absolute value.  In addition, a list of all variables that are used in this paper is given in table 1.

\color{black}
\section{System Model}
\label{sec:SYSTEM MODEL}

\begin{figure}
  \centering
  \includegraphics[width=0.7\textwidth]{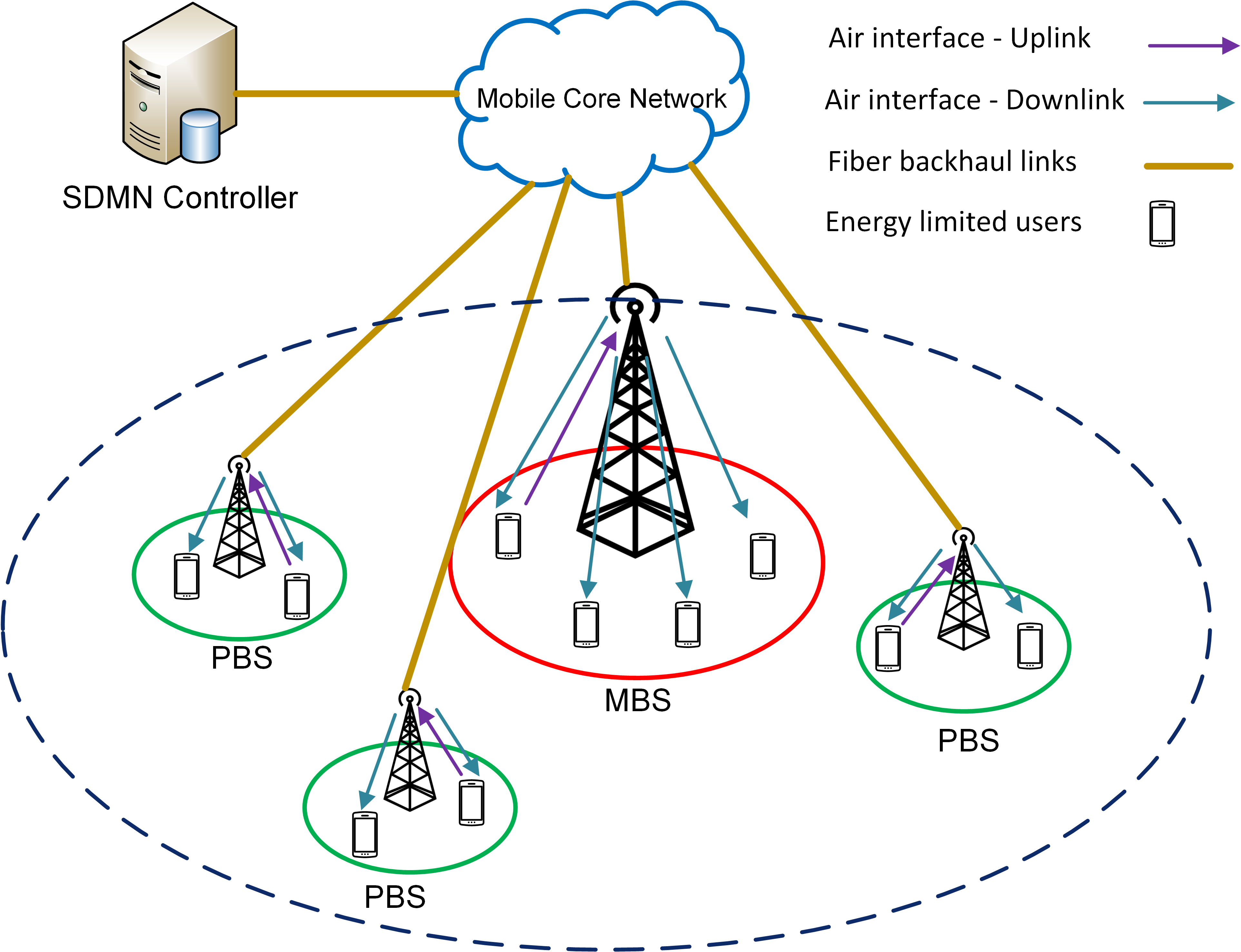}
  \caption{ A Heterogeneous network with energy limited users and SDMN controller. }
  \label{Fig:1}
\end{figure}

As depicted in Figure \ref{Fig:1}, we consider a heterogeneous network consists of one MBS and some PBSs which is a possible candidate for implementing 4G and beyond technologies. In order to assign users to the appropriate BS, we employ user association algorithm which can be implemented in SDMN in practical cases. We consider $K$ single-antenna users and $J$ base stations where there is a single MBS and $(J-1)$ PBSs. The MBS enjoys the Massive MIMO technique and thus is equipped with large-scale antennas and the PBSs exploit the conventional MIMO.
 As in \cite{Xu}, the considered channel model is ${{\bf h}_{j,k}} = {{\bf g}_{j,k}}{l_{j,k}}$, where ${{\bf h}_{j,k}}$ is the vector of channel coefficients between BS $j$ and user $k$. In addition, ${{\bf g}_{j,k}}$ and ${l_{j,k}}$ are the small scale fading coefficient vector and the large scale fading coefficient between BS $j$ and user $k$, respectively. Defining channel matrix ${{\bf H}_j}$ at BS $j$ as ${{\bf H}_j} = [{{\bf h}_{j,1}},...,{{\bf h}_{j,K}}]$, received signals by users connected to BS $j$ can be expressed as

 \begin{eqnarray}\label{1}
{{\bf y}_j} = {{\bf H}_j}{{\bf W}_j}{{\bf d}_j} + {{\bf n}_k},
\end{eqnarray}

\noindent where ${\bf n}_k$ is the zero mean Gaussian noise vector with variance ${\sigma}_k$, ${\bf d}_j$ is data sent from BS $j$ and ${\bf W}_j$ is the precoding matrix. Assuming that each user is capable of harvesting energy from the received signals by PS strategy, $\alpha$ portion of the power of the received signal is considered for data detection and $(1-\alpha)$ portion is considered for energy harvesting. Defining $x_{j,k}$ as association parameter which is assumed $1$ when user $k$ and BS $j$ are connected and $0$ for other conditions.
In this regard, data-rate of user $k$ from BS $j=1$ in the DL can be calculated as \cite{Zhao}

  \begin{eqnarray}\label{2}
R_{j,k}^{DL} =
{x_{j,k}}{\log _2}(1 + \frac{{{M_j} - {L_j} + 1}}{{{L_j}}}\frac{{{\alpha _{j,k}}{P_j}{l_{j,k}}}}{{{{\sigma}_{D_k} ^2}+{\alpha _{j,k}}{{\sigma}_k ^2} + {\alpha _k}\sum\nolimits_{j' \ne j} {{P_{j'}}{l_{j',k}}} }}),
\end{eqnarray}

\noindent where ${{\sigma}_{D_k} ^2}$ is the additional noise power of decoding at user $k$, ${\alpha _{j,k}}$ is the power factor of user $k$ connected to BS $j$,  $P_j$ is the transmit power of BS $j$, $L_j$ is the maximum number of users that can be served by BS $j$ and $M_j$ is the number of antennas at BS $j$. Then, data rate achieved by user $k$ from PBS $j$ when ${x_{j,k}}=1$  can be expressed as

  \begin{eqnarray}\label{3}
R_{j,k}^{DL} = {\log _2}(1 + \frac{{{\alpha _{j,k}} {P_j}|{\bf h}_{j,k}^H{{\bf h}_{j,k}}{|^2}}}{{{{\sigma}_{D_k} ^2}+{\alpha _{j,k}}{{\sigma}_k ^2} + {\alpha _{j,k}}{P_j} \sum\nolimits_{k' \ne k} {|{\bf h}_{j,k'}^H{{\bf h}_{j,k'}}{|^2}} }}),
\end{eqnarray}

\noindent where, ${\bf W}_j$ is considered to be normalized transpose conjugate of the channel matrices which is the optimum value for the precoder. In the DL, the received power at user $k$ from BS $j$ in case of ${x_{j,k}}=1$ is calculated as

 \begin{equation}\label{6}
{RP_{j,k}} = (1 - {\alpha _{j,k}} )trace({P_j}|{\bf h}_{j,k}^H{{\bf h}_{j,k}}{|^2} + \sigma _k^2).
\end{equation}

In the DL, we aim to maximize the sum-rate of the network as well as the received energy from the BSs. In this regard, we introduce a utility function in (\ref{aa}) that contain all our concerns.
 \begin{equation}\label{aa}
\begin{array}{l}
{U_{j,k}}= {\log({RP_{j,k}})+\log({R_{j,k}^{DL}}) } \\

\end{array}
\end{equation}

As it is shown in (\ref{aa}), logarithm function is employed in the utility function since it has the capability of providing fairness. In addition, the utility function appropriately presents the trade-off between the amount of data-rate and the amount of received power.

\section{Optimization Structure in SDMN} \label{sec:3}

 In order to optimize DL, SDMN should apply an appropriate approach to associate users to the optimum BSs, based on the utility function. In the DL association we study the process of optimizing ${\alpha _{j,k}}$ and $x_{j,k}$ and then compare it with the conventional maximum rate user association which is explained later. The user association problem can be summarized as

 \begin{equation}\label{4}
\begin{array}{l}
\mathop {\max }\limits_{{\alpha _{j,k}},{x_{j,k}}} \,\,\sum\limits_{k = 1}^K {\sum\limits_{j = 1}^J {{x_{j,k}}{U_{j,k}}} } \\
s.t.\\
\sum\nolimits_k {{x_{j,k}}}  \le {L_{j\,\,}}\,\,\,j = 1,2,...,J\\
\sum\nolimits_j {{x_{j,k}}}  \le 1\,\,\,\,\,\,k = 1,2,...,K \\
0\leq{\alpha _{j,k}}\leq 1 \\
{{x_{j,k}}}\in \{{0,1}\}.
\end{array}
\end{equation}

In (\ref{4}), the first constraint represents that the number of connected users to BS $j$ is not allowed to exceed its upper bound and the second condition illustrates that each user should be connected to only one BS.
It is observable that the problem (\ref{4}) is separable into two problems: optimizing ${\alpha _{j,k}}$ by each user and choosing the optimal association by the SDMN controller. In this regard, each user $k$ solves the following optimization problem.

 \begin{equation}\label{44}
\begin{array}{l}
\mathop {\max }\limits_{{\alpha _{j,k}}} {{U_{j,k}}} \\
s.t.\,\,\,
0\leq{\alpha _{j,k}}\leq 1
\end{array}
\end{equation}

These $KJ$ optimization problems can be solved in parallel and without any interactions. Then, with known ${\alpha _{j,k}}$ for each user, the optimal values are transmitted to the SDMN controller and the following optimization problem is considered to obtain the association vector.

 \begin{equation}\label{444}
\begin{array}{l}
\mathop {\max }\limits_{{x_{j,k}}} \,\,\sum\limits_{k = 1}^K {\sum\limits_{j = 1}^J {{x_{j,k}}{U_{j,k}}} } \\
s.t.\\
\sum\nolimits_k {{x_{j,k}}}  \le {L_{j\,\,}}\,\,\,j = 1,2,...,J\\
\sum\nolimits_j {{x_{j,k}}}  \le 1\,\,\,\,\,\,k = 1,2,...,K \\
{{x_{j,k}}}\in \{{0,1}\}.
\end{array}
\end{equation}

 As $x_{j,k}$ is an integer parameter, the problem (\ref{444}) is hard to tackle, in this regard we simplify and relax this problem to the following one which $x_{j,k} \in (0,1)$ and it is proved in \cite{Zhao} that these two problems eventuate in the same results.

 \begin{equation}\label{5}
\begin{array}{l}
\mathop {\max }\limits_{{\bf x}} \,{{\bf U}^T} {\bf x}\\
s.t.\\
\sum\nolimits_k {{x_{(j - 1)K + k}}}  \le {L_{j\,\,}}\,\,\,j = 1,2,...,J\\
\sum\nolimits_j {{x_{k + (j - 1)K}}}  \le 1\,\,\,\,\,\,k = 1,2,...,K \\
0\leq{{x_{j,k}}}\leq 1.
\end{array}
\end{equation}

In (\ref{5}), ${\bf x}=[{\bf x}_{11},...,{\bf x}_{K1},...,{\bf x}_{1J},...,{\bf x}_{KJ}]$ and ${\bf U}=[{U}_{11},...,{ U}_{K1},...,{ U}_{1J},...,{ U}_{KJ}]$ is the utility function vector where $U_{j,k}$ is defined as the utility function.
This approach can be compared with its counterpart while considering data-rate instead of utility function \cite{Zhao}. In this regard, the optimization problem eventuate in the data-rate maximization regardless of the amount of harvested energy. In this approach, $U_{j,k}$ is replaced with $R_{j,k}$ which is the corresponding data-rate of the User $k$ served by BS $j$.
 Problem (\ref{5}) is a linear optimization problem and can be solved easily. Association achieved in DL is considered in UL and other communication phases for coherence time of the channel. Hence, considering associated cells in DL, users can harvest energy from DL signals and consume that energy in UL.
  Algorithm 1 is proposed for solving (\ref{5}).

   \makeatletter
\def\BState{\State\hskip-\ALG@thistlm}
\makeatother
\begin{algorithm}
\caption{: DL users association and power factor optimization}\label{etabisection}
\begin{spacing}{1.2}
\begin{algorithmic}[1]
\State $\textbf{Initialize:} \ t,\;P_j, \; M_j, \; L_j, \; {\bf H}_j, j=1,...,J $
\State  solve KJ parallel optimization problems with respect to ${\alpha _{j,k}}$.
\State  with known ${\alpha _{j,k}}$, solve (\ref{5}) to find the optimized ${\bf x}$.
\end{algorithmic}
\end{spacing}
\end{algorithm}

   In order to eliminate interference in UL, at each time epoch at each cell only one user is allowed to transmit data. Consequently, choosing the best user and its transmit power leads to another optimization problem. In this study, we consider the finite battery capacity of users and hence after battery full charge, the received energy should stay in energy queue. Lyapunov function describes queue features of energy appropriately and simplifies the optimization problem. In UL, the data-rate is

 \begin{equation}\label{7}
R_{j,k}^{UL}(t) =  {{{\log }_2}(1 + \frac{{{\lambda_{j,k}}^2{P_{k}}(t)}}{{\sigma_j ^2}})} ,
\end{equation}
where  $\sigma_j^2$ is the white Gaussian noise power at BS $j$ and ${\lambda_{j,k}}$ is the  nonzero eigenvalue of the channel between BS $j$ and user $k$ and $P_{k}$ is the average power of each user. Assuming $C_k$ as the maximum battery capacity of user $k$ and harvested power at user $k$ by $S_k$ we have

 \begin{equation}\label{8}
{S_k}(t + 1) = \min \{ {S_k}(t) - {P_k}(t) + {RP_{j,k}(t)},{C_k}\}.
\end{equation}

Due to the limited battery capacity of users, we have ${P_k}(t) \le {S_k}(t)$. Deriving queue theory for harvested energy, Lyapunov function can be defined as $L({S_k}(t)) = \frac{1}{2} {S_k^2} (t)$. In this regard, drift-plus-penalty (DPP) can be expressed as

 \begin{equation}\label{9}
DPP =  - R_{j,k}^{UL}(t) + \frac{1}{2}{V_k}(t)(S_k^2(t + 1) - S_k^2(t)),
\end{equation}

\noindent where $V_k(t)$ is an appropriate Lyapunov weight for each user. Then

 \begin{equation}\label{10}
 \begin{array}{l}
\Delta ({S_k}(t)) = L({S_k}(t + 1)) - L({S_k}(t))\le B + \sum\limits_{k = 1}^K {{V_k}(t){S_k}(t)({RP_{j,k}}(t) - {P_k}(t))},
\end{array}
\end{equation}

\noindent where $B$ is a constant that $B \ge \frac{1}{2}\sum\limits_{k = 1}^K {{V_k}} (t)(RP_{j,k}^2(t) + P_k^2(t))$.

\textit{ Proof:} Plugging ${S_k}(t + 1)$ in $\Delta ({S_k}(t))$ and assuming ${S_k}(t) - {P_k}(t) + {RP_{j,k}(t)}\leq{C_k}$ we have

  \begin{equation}\label{proof}
 \begin{array}{l}
\Delta ({S_k}(t)) = L({S_k}(t + 1)) - L({S_k}(t))
=\frac{1}{2}\sum\limits_{n = 1}^K {S_k^2} (t+1) -\frac{1}{2}\sum\limits_{n = 1}^K {S_k^2} (t)\\
=\frac{1}{2}\sum\limits_{n = 1}^K \{{S_k^2} (t)+{P_k^2}(t)-2 {S_k^2} (t){P_k^2}(t)
 +{RP_{j,k}^2}(t)+2{RP_{j,k}}(t)({S_k} (t)-{P_k}(t))-{S_k^2} (t)\}\\
 =\frac{1}{2}\sum\limits_{k = 1}^K {{P_k^2}(t)+{RP_{j,k}^2}(t)-2{RP_{j,k}}(t){P_k} (t)}
+ \sum\limits_{k = 1}^K {{S_k}(t)({RP_{j,k}}(t) - {P_k}(t))}.
\end{array}
\end{equation}

 In case of ${S_k}(t) - {P_k}(t) + {RP_{j,k}(t)}>{C_k}$, drift is calculated as
   \begin{equation}\label{proof2}
 \begin{array}{l}
\Delta ({S_k}(t)) = L({S_k}(t + 1)) - L({S_k}(t))
=\frac{1}{2}\sum\limits_{n = 1}^K {S_k^2} (t+1) -\frac{1}{2}\sum\limits_{n = 1}^K {S_k^2} (t)
=\frac{1}{2}\sum\limits_{n = 1}^K \{{C_k^2}-{S_k^2} (t)\}.
\end{array}
\end{equation}

It should be noticed that as the ultimate goal is to minimize the drift, (\ref{proof}) should be considered for the DPP optimization.
 Then, we define a control parameter for each user and each time epoch as $V_k(t)$ in order to control the trade-off between the penalty and the drift terms in the DPP. In this regard, DPP is formulated as

 \begin{equation}\label{11}
DPP \le B - \sum\limits_{k = 1}^K {R_{j,k}^{UL}} (t) + \sum\limits_{k = 1}^K {{V_k}(t){S_k}(t)({RP_{j,k}}(t) - {P_k}(t))},
\end{equation}
\noindent where
\begin{equation*}\label{BB}
 \begin{array}{l}
B=\frac{1}{2}\sum\limits_{k = 1}^K {{{V_k}} (t){P_k^2}(t)+{RP_{j,k}^2}(t)-2{RP_{j,k}}(t){P_k} (t)}
\ge \frac{1}{2}\sum\limits_{k = 1}^K {{V_k}} (t)(RP_{j,k}^2(t) + P_k^2(t))
\end{array}
\end{equation*}
is constant.
Consequently, the only term that should be optimized is $ - \sum\limits_{k = 1}^K {R_{j,k}^{UL}} (t) + \sum\limits_{k = 1}^K {{V_k}{S_k}(t)({RP_{j,k}}(t) - {P_k}(t))} $.
Therefore, the DPP optimization problem that should be solved in SDMN is summarized as

 \begin{equation}\label{12}
\begin{array}{l}
\mathop {\max {\mkern 1mu} {\mkern 1mu} }\limits_{P_k(t), k} \{ \sum\limits_{k = 1}^K {R_{j,k}^{UL}} (t) + \sum\limits_{k = 1}^K {{V_k}(t){S_k}(t){P_k}(t)} \} \\
\,\,\,\,\,\,\,\,\,s.t.{\mkern 1mu} {\mkern 1mu} {\mkern 1mu} {\mkern 1mu} {\mkern 1mu} {\mkern 1mu} {\mkern 1mu} {\mkern 1mu} 0 \le {P_k}(t) \le {S_k}(t){\mkern 1mu} {\mkern 1mu} {\mkern 1mu} {\mkern 1mu} {\mkern 1mu} {\mkern 1mu} {\mkern 1mu} {\mkern 1mu} {\mkern 1mu} {\mkern 1mu} {\mkern 1mu} {\mkern 1mu} \\
\,\,\,\,\,\,\,\,{\mkern 1mu} Only\,{\mkern 1mu} one\,{\mkern 1mu} active\,{\mkern 1mu} user{\mkern 1mu} per\,{\mkern 1mu} cell.
\end{array}
\end{equation}

It is noteworthy to mention that the optimization problem in (\ref{12}) should be held for $J$ BSs in the SDMN and each BS is allowed to choose one active user for each UL time epoch. In order to deal with the optimization problem regarding to (\ref{12}), similar to the adopted approach for solving (\ref{5}), we can divide the optimization problem into two independent ones. Consequently, first we find the optimal transmit power for each user at each cell then regarding to the optimum values we derive the optimization problem for each cell in order to find the best user for obtaining the UL air interface.
The following optimization problem should be solved at each user.
\begin{equation}\label{13}
\begin{array}{l}
\mathop {\max {\mkern 1mu} {\mkern 1mu} }\limits_{P_k(t)} \{ {R_{j,k}^{UL}} (t) + {{V_k}(t){S_k}(t){P_k}(t)} \} \\
\,\,\,\,\,\,\,\,\,s.t.{\mkern 1mu} {\mkern 1mu} {\mkern 1mu} {\mkern 1mu} {\mkern 1mu} {\mkern 1mu} {\mkern 1mu} {\mkern 1mu} 0 \le {P_k}(t) \le {S_k}(t){\mkern 1mu} {\mkern 1mu} {\mkern 1mu} {\mkern 1mu} {\mkern 1mu} {\mkern 1mu} {\mkern 1mu} {\mkern 1mu} {\mkern 1mu} {\mkern 1mu} {\mkern 1mu} {\mkern 1mu} \\
\,\,\,\,\,\,\,\,{\mkern 1mu} Only\,{\mkern 1mu} one\,{\mkern 1mu} active\,{\mkern 1mu} user{\mkern 1mu} per\,{\mkern 1mu} cell.
\end{array}
\end{equation}

Then, with the known optimum $P_k(t)$ of each user, the total optimization problem in SDMN is derived in order to find the user at each cell which is allowed to utilize the air interface. Thus, the optimization problem is

\begin{equation}\label{14}
\begin{array}{l}
\mathop {\max {\mkern 1mu} {\mkern 1mu} }\limits_{k} \{ \sum\limits_{k = 1}^K {R_{j,k}^{UL}} (t) + \sum\limits_{k = 1}^K {{V_k}(t){S_k}(t){P_k}(t)} \} \\
\,\,\,\,\,\,\,\,\,s.t.{\mkern 1mu} {\mkern 1mu} {\mkern 1mu} {\mkern 1mu} {\mkern 1mu} {\mkern 1mu} {\mkern 1mu} {\mkern 1mu} 0 \le {P_k}(t) \le {S_k}(t){\mkern 1mu} {\mkern 1mu} {\mkern 1mu} {\mkern 1mu} {\mkern 1mu} {\mkern 1mu} {\mkern 1mu} {\mkern 1mu} {\mkern 1mu} {\mkern 1mu} {\mkern 1mu} {\mkern 1mu} \\
\,\,\,\,\,\,\,\,{\mkern 1mu} Only\,{\mkern 1mu} one\,{\mkern 1mu} active\,{\mkern 1mu} user{\mkern 1mu} per\,{\mkern 1mu} cell.
\end{array}
\end{equation}

\section{Control Parameter Calculation}
\label{sec:control parameter}
In order to provide fairness among users and construct an appropriate trade-off between penalty and energy drift, choosing the proper value for $V_k(t)$ is of the great importance. In this regard, we introduce a dynamic determination of the control parameter based on the UL data-rate of each user. It should be noted that without dynamic approach, when the data-rate of user is high enough (due to the high channel quality in case of low distances), the optimization problem tend to allocate the air interface to the nearest user. While adopting the dynamic control parameter, when the data-rate term in the objective function is low, the control parameter should be high enough in order to intensify the effect of cumulated energy to maintain fairness. Consequently, the control parameter for each user at each time epoch can be defined as
\begin{equation}\label{control}
{V_k}(t) = \max \{ \widetilde {{R_k}}(t)\}  - \widetilde {{R_k}}(t)
\end{equation}
\noindent where
$\widetilde {{R_k}}(t) = \sum\limits_{\tau  = t - t'}^{t - 1} {{R_k}(\tau )} $
is the sum-rate of user $k$ in the interval $(t - t',t-1)$. In this regard, the control factor is $0$ at the beginning of the process and changes over the time in order to prevent the SDMN to neglect the effect of energy term. Algorithm 2 represents the UL scheduling paradigm of the proposed scheme in one time epoch and Algorithm 3 summarizes the whole optimization procedure in SDMN.

   \makeatletter
\def\BState{\State\hskip-\ALG@thistlm}
\makeatother
\begin{algorithm}
\caption{: UL user association and energy scheduling}\label{etabisection}
\begin{spacing}{1.2}
\begin{algorithmic}[1]
\State $\textbf{Initialize:} \ j=1,\;P_j, \; M_j, \; L_j, \; {\bf H}_j $
\While {($j\leq J$)}
\State  Calculate $V_k(t)$ using (\ref{control});
\State  Solve (\ref{13}) to obtain $P_k(t)$;
\State  Controller solves (\ref{14}) to dedicate air interface to the optimum user;
\State  j=j+1;
\EndWhile \textbf{end while}
\end{algorithmic}
\end{spacing}
\end{algorithm}

   \makeatletter
\def\BState{\State\hskip-\ALG@thistlm}
\makeatother
\begin{algorithm}
\caption{: SDMN controller optimization}\label{etabisection}
\begin{spacing}{1.2}
\begin{algorithmic}[1]
\State $\textbf{Initialize:} \ t=1,\ t_0=1,\;P_j, \; M_j, \; L_j, \; {\bf H}_j, j=1,...,J $
\While {($t\leq t_{end}$)}
\If {($(t-t_0+1)==t_{Coh}$ or $(t-t_0+1)==1$)} $\;Use \  Algorithm \  1 \  and \  t_0=t;$
\EndIf
\State  Use Algorithm 2;
\State  $t=t+1$;

\EndWhile \textbf{end while}
\end{algorithmic}
\end{spacing}
\end{algorithm}
\section{Simulation Results} \label{sec:SIMULATION}

\begin{figure}[!h]
\centering
  \includegraphics[width=1\textwidth,trim={0 5cm 0 5cm},clip]{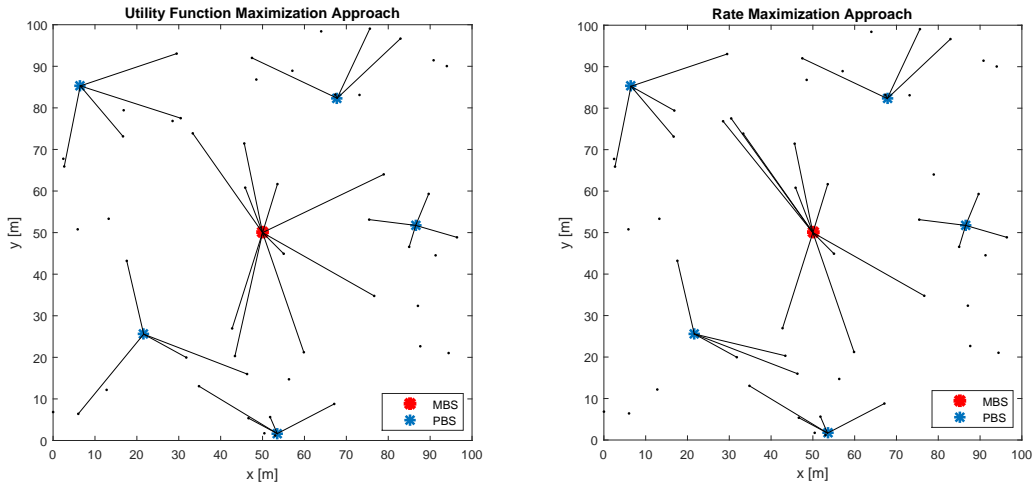}
  \caption{User Association in HetNets.}
  \label{association}
\end{figure}

 In this simulations, we assume $J=6$ and $K=50$ and the number of MBS antennas is set as $M=100$ and the number of PBS antennas is 4. In addition, the maximum number of users that the base stations can serve are 10 and 4 for the MBS and PBSs, respectively. Channels are modeled as complex Gaussian zero-mean unit-variance random variables. We assume $l_{j,k}=1/(1+(d_{j,k}/40)^{3.5})$ for the path loss between user $k$ and BS $j=1$ and $l_{j,k}=1/(1+(d_{j,k}/40)^{4})$ for the path loss between user $k$ and BS $j, j\neq 1$, where $d_{j,k}$ is the distance between user $k$ and BS $j$ \cite{Xu}. All noise powers are assumed to be 0.2 and we assume $C_k=300$ and $t'=5$. Figure \ref{association} exhibits the user association in a HetNet based on the utility function maximization and sum-rate maximization approaches. It should be noted that utilizing different strategies results is different system architectures.

\begin{figure}[!h]
\centering
  \includegraphics[width=0.7\textwidth]{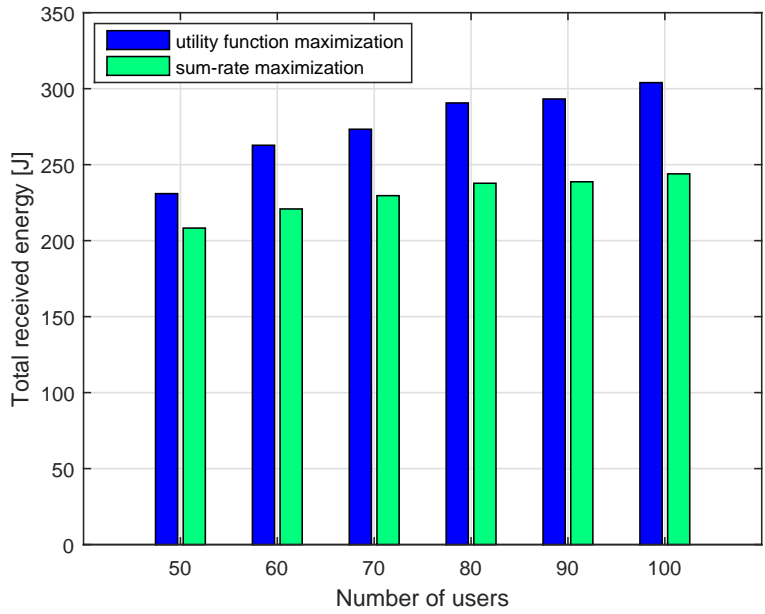}
  \caption{Total received energy versus number of users for two different approaches.}
  \label{bar1}
\end{figure}

Figure \ref{bar1} represents the total received energy versus number of users for utility function maximization and sum-rate maximization approaches. It should be noticed that the enhancement in the number of users provides more freedom in the optimization problem and leads to an increase in the amount of received energy. Moreover, by exploiting the introduced utility function, which considers the trade-off between the energy and sum-rate, we face the received energy increase.

\begin{figure}[!h]
\centering
  \includegraphics[width=0.7\textwidth]{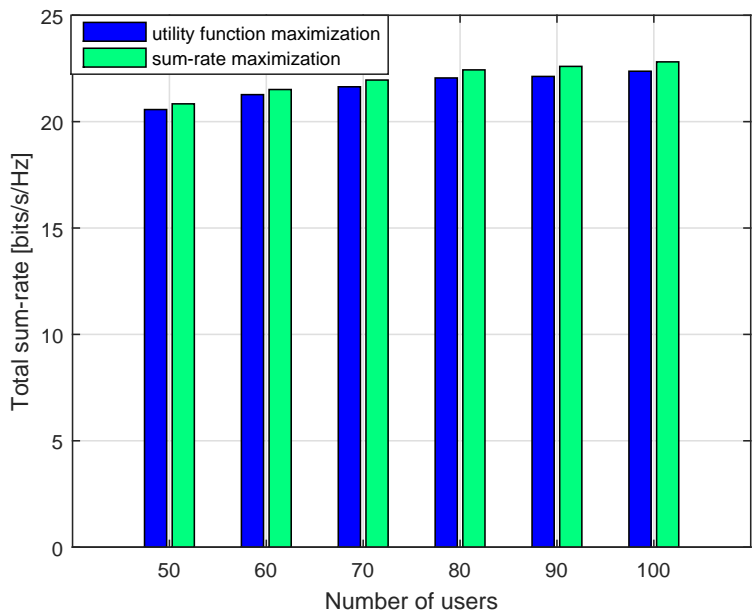}
  \caption{Total sum-rate versus number of users for two different approaches.}
  \label{bar2}
\end{figure}

\begin{figure}[!h]
\centering
  \includegraphics[width=0.7\textwidth]{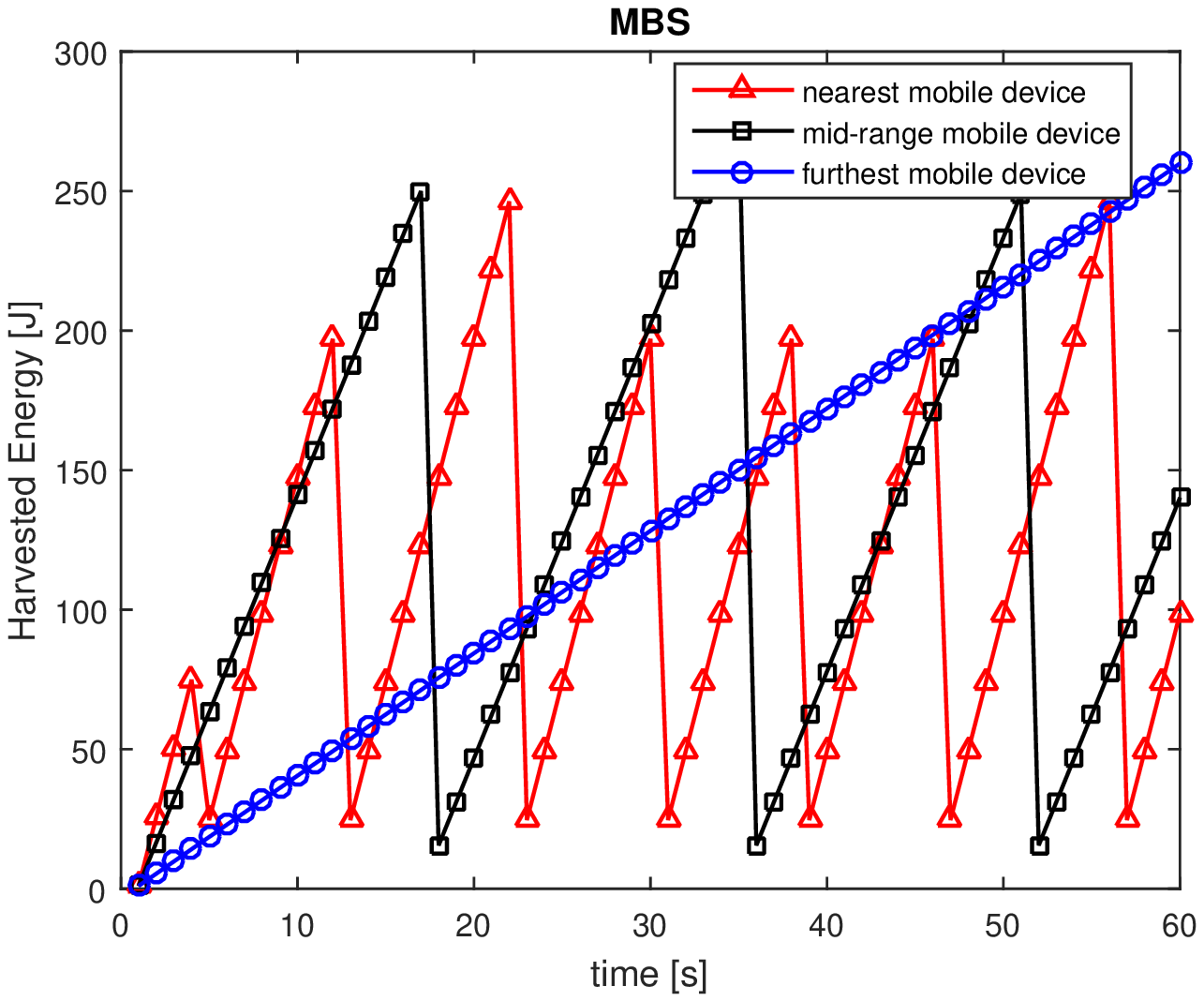}
  \caption{Comparison of Harvested Energy for MBS Users located in different distances and based on maximum rate approach.}
  \label{MBS1}
\end{figure}

\begin{figure}[!h]
\centering
  \includegraphics[width=0.7\textwidth]{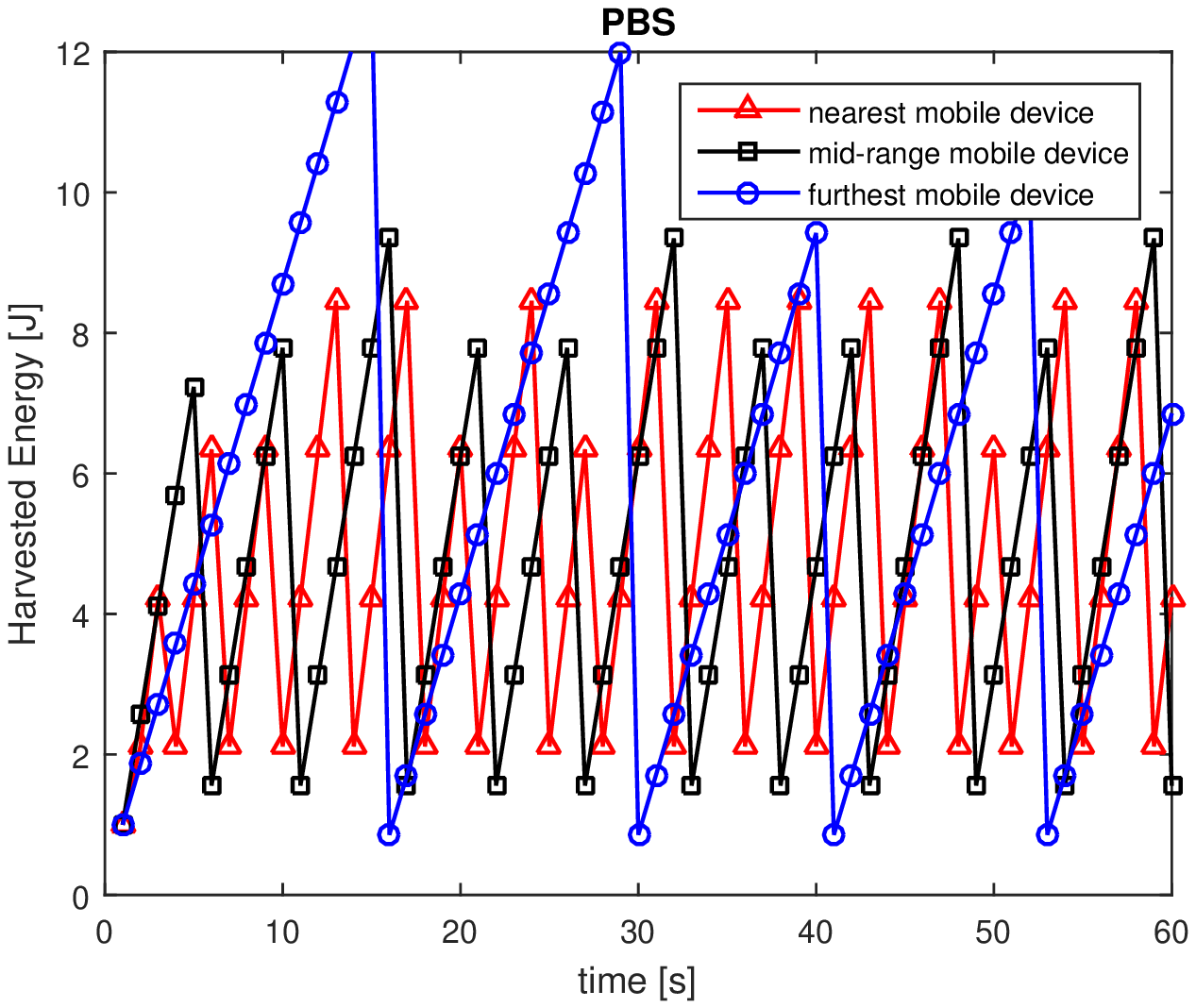}
  \caption{Comparison of Harvested Energy for PBS Users located in different distances and based on maximum rate approach.}
  \label{PBS1}
\end{figure}

\begin{figure}[!h]
\centering
  \includegraphics[width=0.7\textwidth]{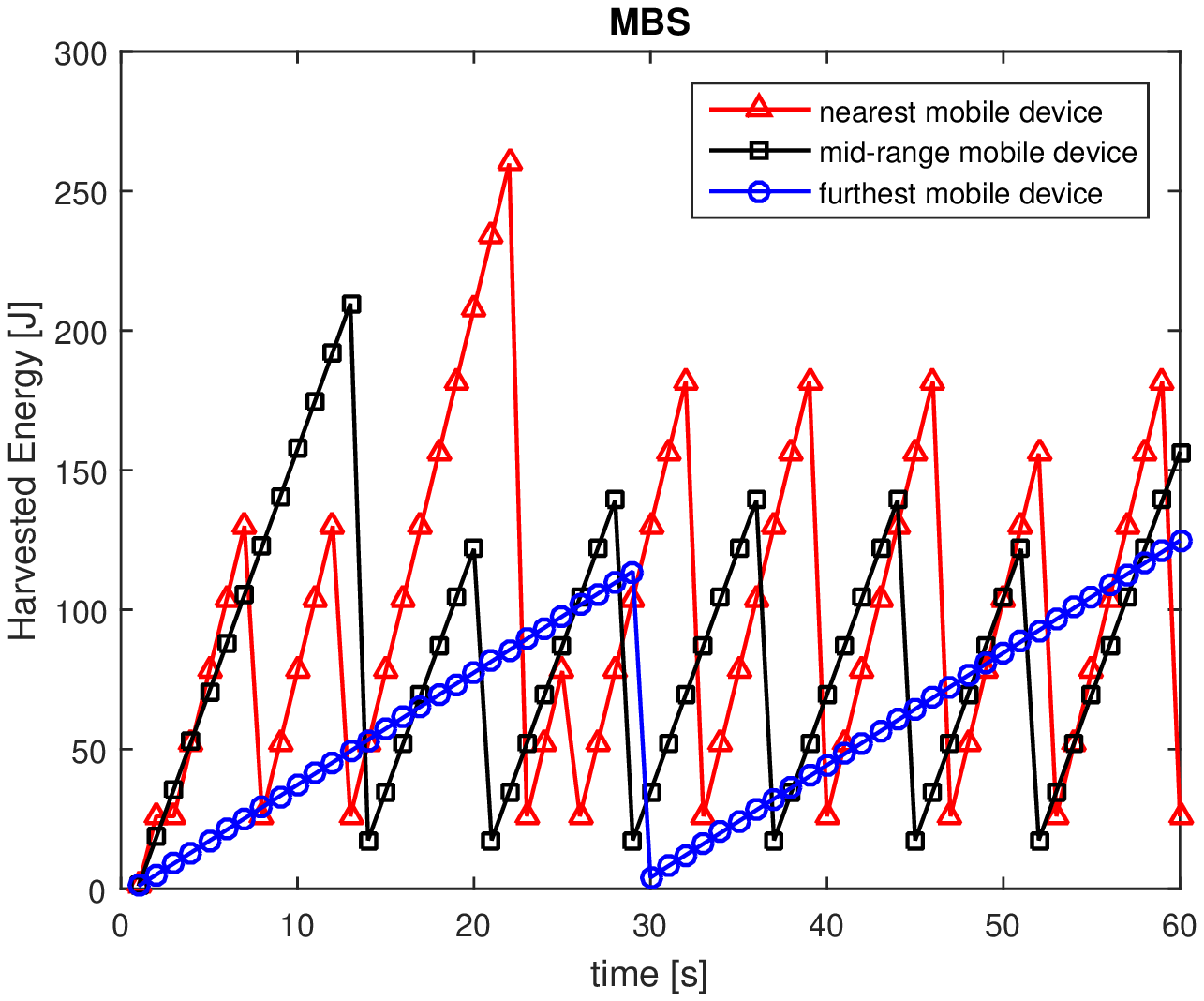}
  \caption{Comparison of Harvested Energy for MBS Users located in different distances and based on Lyapunov approach.}
  \label{MBS2}
\end{figure}

\begin{figure}[!h]
\centering
  \includegraphics[width=0.7\textwidth]{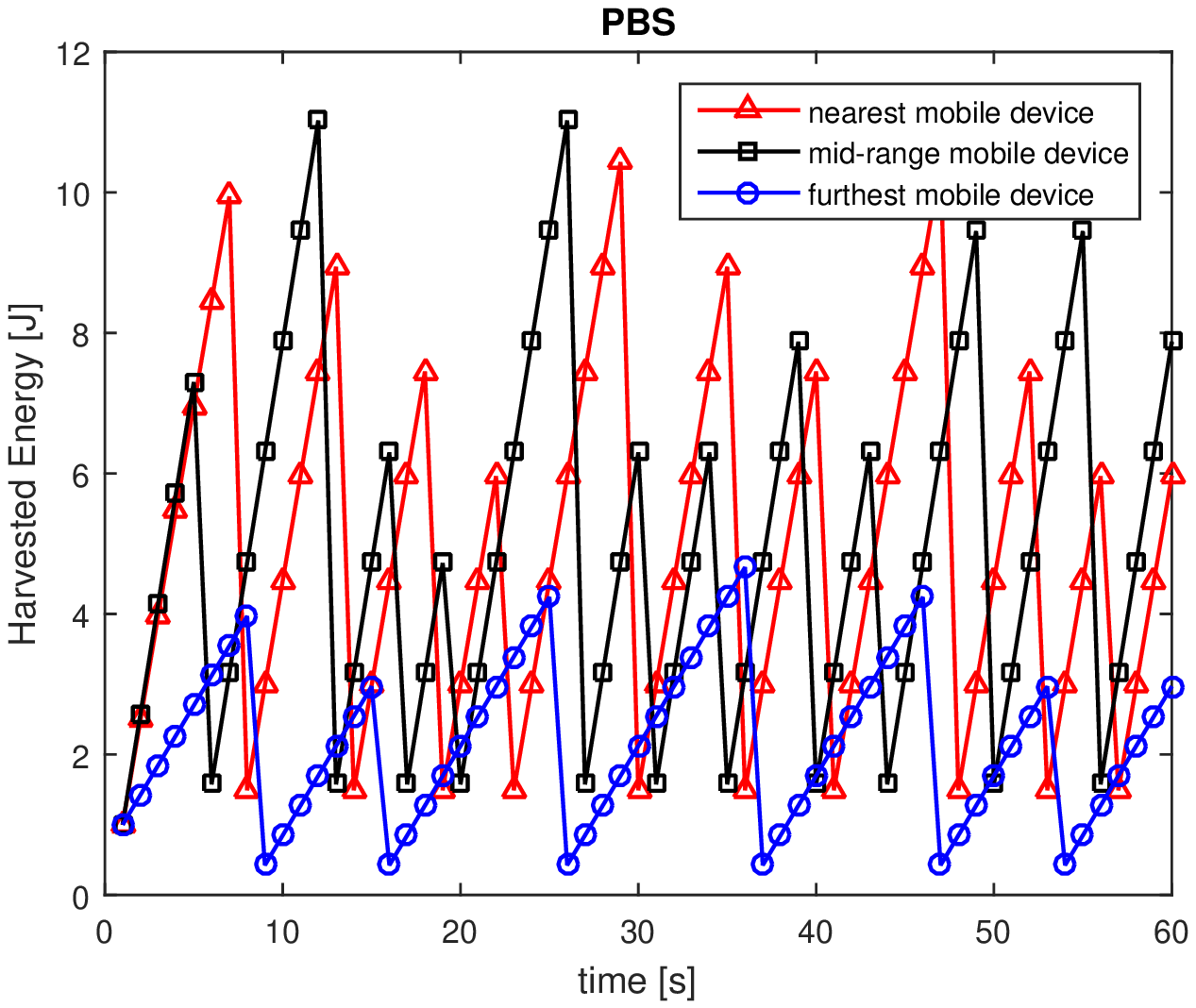}
  \caption{Comparison of Harvested Energy for PBS Users located in different distances and based on Lyapunov approach.}
  \label{PBS2}
\end{figure}

Figure \ref{bar2} represents the total sum-rate versus number of users for the two approaches. In this figure, it is obvious that the sum-rate maximization method results in higher sum-rates. However, the deviation between two approaches are low enough that the weakness of the utility-function maximization approach in the sum-rate can be overlooked in comparison with its energy benefits.

Figure \ref{MBS1} expresses the harvested energy at users' batteries in macro cell, based on the maximum rate approach. The decrease in the energy level is an indicator of participation in the UL communications by users. It can be seen that the nearest user benefits from this approach as its discharge frequency is higher than other users. It is noteworthy to say that considering maximum rate approach provides users with less fairness in the network.

Figure \ref{PBS1} expresses the harvested energy at users' batteries in pico cell, based on the maximum rate approach. As it can be seen, the frequency of discharging is not the same among users which represents low-level of fairness. However, due to the less coverage of the PBS, the level of unfairness among users is less than the MBS.

Figure \ref{MBS2} represents the harvested energy at users' batteries in MBS based on the Lyapunov approach which considers the trade-off between the network data-rate and the battery levels. It should be noticed that employing Lyapunov approach provides users with more fairness, since the discharge frequency of the further users is increased.

Figure \ref{PBS2} shows the harvested energy at users' batteries in a pico cell considering the Lyapunov function in optimization. It can be seen that the discharge frequency of users has become closer to each other, albeit there still exist frequency differences between them. This variation stems from the fact that the further users receive less power levels in the DL and hence they enjoy less available transmit power which weakens their probability of obtaining the UL air interface. It should be noted that in the case of providing users with random power levels independently, the frequency of discharge will be almost the same among users which represents the fairness among users that is fulfilled by employing the lyapunov approach. However, in this paper in order to exploit the received power in DL efficiently, we utilize the users' received power in DL for effective communications in UL and hence the further users suffer from low received power which we tried to compensate via the proposed approach.

\section{Conclusion} \label{6}

In this paper, the UL and DL optimization in SDMN has been investigated while the users are energy limited and harvest energy for dealing with this problem. In the DL, user association problem has been derived based on the problem of utility function maximization in order to optimize cell architecture as well as harvesting factors of users. This utility function has been defined as the sum of logarithm of data-rate and harvested energy. Then, adopting the same cell architecture for the UL in the coherence time of the channels, the user energy scheduling has been proposed in order to control the dedication of air interface via UL and its transmit power. In this regard, Lyapunov drift-plus-penalty approach has been employed with dynamic control parameter. This network optimization has led to a better user association in terms of data-rate and harvested energy and has provided better fairness and battery stabilization among energy harvesting users. In addition, as wireless sensors and IoT devices are mostly power limited, energy scheduling is a critical issue in IoT in 5G mobile networks. Consequently, energy scheduling and resource allocation can be a potential future work in 5G SDMNs while considering mobile users, wireless sensors and IoT devices jointly.


\end{document}